\documentstyle[manuscript,aps,epsfig]{revtex}
\begin{document}

\topmargin 20pt

\newcommand{\be}{\begin{equation}}
\newcommand{\ee}{\end{equation}}
\newcommand{\bea}{\begin{eqnarray}}
\newcommand{\eea}{\end{eqnarray}}

\newcommand{\0}{\bar{0}}
\newcommand{\1}{\bar{1}}
\newcommand{\2}{\bar{2}}
\newcommand{\3}{\bar{3}}

\newcommand{\ta}{\tilde{a}}
\newcommand{\td}{\tilde{d}}
\newcommand{\tV}{\tilde{V}}


\title{Kasner Asymptotics of Mixmaster Ho\v{r}ava-Witten Cosmology.}
\author{Mariusz P. D\c{a}browski
\footnote{E-mail:mpdabfz@uoo.univ.szczecin.pl}\\
Institute of Physics, University of Szczecin, Wielkopolska 15,
          70-451 Szczecin, Poland}
\date{\today}

\maketitle

\begin{abstract}

Bianchi type I and type IX ('Mixmaster') geometries are investigated
within the framework of Ho\v{r}ava-Witten cosmology. We consider
the models for which the fifth coordinate is a $S^1/Z_2$ orbifold
while the four coordinates are such that the 3-space is
homogeneous and has geometry of Bianchi type I or IX while the rest six
dimensions have already been compactified on a Calabi-Yau space. In
particular, we study Kasner-type solutions of the Bianchi I field equations and discuss Kasner
asymptotics of Bianchi IX field equations. We are able to recover the isotropic 3-space
solutions found by Lukas {\it et al} \cite{lukas}. Finally, we discuss if
such Bianchi IX configuration can result in chaotic behaviour of these Ho\v{r}ava-Witten
cosmologies.

\end{abstract}


PACS number(s): 98.80.Hw, 04.50.+h, 11.25.Mj, 98.80.Cq


\newpage

String cosmology has attracted a lot of interest recently (for a review see \cite{superjim}),
especially in the context of duality symmetry, which is a striking feature of the underlying
string theory and justifies a kinetic-energy-driven inflation, known as pre-big-bang
inflation \cite{superinf}. Pre-big-bang inflation is the result of the admission of the
cosmological solutions for bosonic low-energy-effective-action for strings \cite{fradcall}.
Bosonic action is the simplest
stringy action and in view of duality symmetry this action together with other
superstring actions are not necessarily the right description of physics at strong coupling
- M-theory. Ho\v{r}ava and Witten \cite{hw} proposed that the right
candidate for M-theory is strongly coupled limit of $E_8 \times E_8$
heterotic superstring theory compactified on a $S^1/Z_2$ orbifold
with $E_8$ gauge fields on each orbifold fixed plane. This means
the gauge fields live on 10-dimensional planes while gravity can
propagate in the whole 11-dimensional bulk. The idea of having
extra dimensions in which only gravity can propagate has been
under intensive studies recently and different scenarios
(including extra time dimensions \cite{extratime}) have been considered
\cite{dimo,kalo,extradim,kam}.

In this paper we will consider the models in which M-theory is
compactified on an orbifold and then reduced to four dimensions
using Calabi-Yau manifold \cite{lukas,reall}. Since the size of the orbifold is
bigger than the size of the Calabi-Yau space then there was a
period in the history of the universe during which the universe
was five-dimensional. That means we can consider the cosmological models
for which the fifth coordinate is an orbifold while the remaining four
coordinates are such that the three-space is homogeneous with
Bianchi type I or IX geometry. The main objective is to study
Kasner-type solutions of Bianchi type I field equations and Kasner asymptotic states
(as a result of Kasner-to-Kasner transitions) of Bianchi type IX field equations.
The form of these solutions allows us to find out whether there is
a possibility for chaotic behaviour in these Ho\v{r}ava-Witten
cosmologies. Similar question was addressed in \cite{bd2} for
pre-big-bang cosmology with the answer that only finite number of
chaotic oscillations are possible.

It is well known that the vacuum BIX homogeneous cosmology in
general relativity is chaotic \cite{bkl}. An infinite
number of oscillations of the orthogonal scale factors occurs in general on any finite
interval of proper time including the singularity at $t=0.$

If a minimally coupled, massless scalar field (e.g. the inflaton)
is admitted, the situation changes. Only a finite number of spacetime
oscillations can occur before the evolution is changed into a state in
which all directions shrink monotonically to zero as the curvature
singularity is reached and the oscillatory behaviour ceases \cite{belkhal}.
This is also the case in 4-dimensional pre-big-bang cosmology
where the role of a scalar field is played by the dilaton
\cite{bd2}. On the other hand, 5-dimensional vacuum Einstein solutions
of Bianchi type IX do not allow chaos to occur either \cite{john85,ishihara,halp}.
The point is that the fifth dimension plays effectively the role
of a scalar field in scalar field cosmologies and stops chaotic
oscillations. In 5-dimensional Ho\v{r}ava-Witten cosmology the
situation is in some ways analogous to both of the above cases which gives some
new interesting points to be made and this is the task of our
paper.

The Ho\v{r}ava-Witten field equations are given by \cite{hw,lukas,reall}

\begin{eqnarray}
R_\mu ^\nu = \nabla _\mu \nabla ^\nu \phi + \frac{\alpha_0^2}{6}
g_{\mu}^{\nu} e^{-2\sqrt{2}\phi} + \sqrt{2}\alpha_0
e^{-\sqrt{2}\phi} \sqrt{\frac{\tilde{g}}{g}} \tilde{g}^{ij} \left[
g_{i\mu} g_j^{\nu} - \frac{1}{2} g_{i\sigma} g_{j}^{\sigma}
\right] \left[ \delta(y) - \delta(y - \pi \lambda) \right] \\
\frac{1}{\sqrt{-g}} \partial_\mu \left( \sqrt{-g} \partial^{\mu} \phi \right)
= - \frac{\sqrt{2}}{3} \alpha_0^2 e^{-2\sqrt{2} \phi} + 2 \alpha
\sqrt{\frac{\tilde{g}}{g}} e^{-\sqrt{2}\phi} \left[ \delta(y) - \delta(y - \pi \lambda)
\right],
\label{eom}
\end{eqnarray}
where $\phi = 1//\sqrt{2} \ln{V}$ and $V$ is a scalar field measuring the deformation of
the Calabi-Yau space, $g_{\mu\nu}$ is the five dimensional metric
tensor while $g_{ij}$ is the four dimensional metric which denotes
the pull-back of the metric on five-dimensional manifold $M_5$ onto the
orbifold fixed four-dimensional manifolds $M^{(1)}_4$ and $M^{(2)}_4$.
In (1)-(2) we have neglected the terms which come from the
three-form on the Calabi-Yau space. Actually, they will not
make any qualitative change in our discussion - this is on the same footing
as it was the case in pre-big-bang models \cite{bd2}. In (1)-(2) $y \in [-\pi \lambda, \pi \lambda]$ is a
coordinate in the orbifold direction and the orbifold fixed planes are at $y = 0, \pi
\lambda$. $Z_2$ acts on $S^1$ by $y \to -y$. The terms involving delta functions
arise from the stress energy on the boundary planes.

Following \cite{lukas} we consider cosmological models of the form
\be
ds_5^2 = - N^2(\tau,y) d\tau^2 + ds_3^2 + d^2(\tau,y) dy^2 ,
\ee
\label{metric}
where
\begin{equation}
ds_3^2=a^2(\tau,y)(\sigma ^1)^2+b^2(\tau,y)(\sigma ^2)^2+c^2(\tau,y)(\sigma ^3)^2,
\end{equation}
is a homogeneous Bianchi type IX 3-metric and the orthonormal forms
$\sigma ^1,\sigma ^2,\sigma ^3$ are given by
\begin{eqnarray}
\sigma ^1 &=&\cos {\psi }d\theta +\sin {\psi }\sin {\theta }d\varphi ,
\\
\sigma ^2 &=&\sin {\psi }d\theta -\cos {\psi }\sin {\theta }d\varphi ,
\\
\sigma ^3 &=&d\psi +\cos {\theta }d\varphi ,
\end{eqnarray}
and the angular coordinates $\psi ,\theta ,\varphi $ span the
following ranges,
\begin{equation}
0\leq \psi \leq 4\pi ,\hspace{0.5cm}0\leq \theta \leq \pi
,\hspace{0.5cm}%
0\leq \varphi \leq 2\pi .
\end{equation}

Similarly as in \cite{lukas,reall} we will look for separable
solutions of the form
\bea
N(\tau,y) &=& n(\tau) \tilde{a}(y) , \nonumber \\
a(\tau,y) &=& \alpha(\tau) \tilde{a}(y) , \nonumber \\
b(\tau,y) &=& \beta(\tau) \tilde{a}(y) , \nonumber \\
c(\tau,y) &=& \gamma(\tau) \tilde{a}(y) ,  \\
d(\tau,y) &=& \delta(\tau) \tilde{a}(y) , \nonumber \\
V(\tau,y) &=& \varepsilon(\tau) \tilde{a}(y) , \nonumber .
\eea

The nonzero components of the field equations read as
(an overdot means a derivative with respect to time $\tau$ and a prime
means a derivative with respect to an orbifold coordinate $y$)

\begin{eqnarray}
\frac{\ta^2}{\td^2} \left[ - \frac{\ta''}{\ta} - \frac{\ta'}{\ta} \left(3\frac{\ta'}{\ta}
- \frac{\ta'}{\ta} \right) - \frac{1}{6} \alpha_0^2 \frac{\delta^2}{\varepsilon^2} \frac{\td^2}{\tV^2}
- \sqrt{2} \alpha_0 \frac{\delta}{\varepsilon} \frac{\td}{\tV} \left( \delta(y) - \delta(y - \pi \lambda)
\right) \right] \nonumber \\
= \frac{\delta^2}{n^2} \left[\frac{\dot{n}}{n}
\left(\frac{\dot{\alpha}}{\alpha} + \frac{\dot{\beta}}{\beta} +
\frac{\dot{\gamma}}{\gamma} + \frac{\dot{\delta}}{\delta} \right)
- \frac{\ddot \alpha}\alpha - \frac{\ddot \beta}\beta - \frac{\ddot \gamma}\gamma
- \frac{\ddot \delta}\delta  - \frac{1}{2} \frac{\dot{\varepsilon}^2}{\varepsilon^2}
\right]  \\
\frac{\ta^2}{\td^2} \left[ - \frac{\ta''}{\ta} - \frac{\ta'}{\ta} \left(3\frac{\ta'}{\ta}
- \frac{\ta'}{\ta} \right) - \frac{1}{6} \alpha_0^2 \frac{\delta^2}{\varepsilon^2} \frac{\td^2}{\tV^2}
- \sqrt{2} \alpha_0 \frac{\delta}{\varepsilon} \frac{\td}{\tV} \left( \delta(y) - \delta(y - \pi \lambda)
\right) \right] \nonumber \\
= \frac{\delta^2}{n^2} \left[ \frac{\dot{\alpha}}{\alpha} \left( \frac{\dot{n}}{n} -
\frac{\dot{\beta}}{\beta} - \frac{\dot{\gamma}}{\gamma} - \frac{\dot{\delta}}{\delta} \right)
- \frac{\ddot \alpha}\alpha -\frac{n^2}{2\alpha^2\beta^2\gamma^2}\left( \left(
\beta^2-\gamma^2\right)^2-\alpha^4\right) \right] , \\
\frac{\ta^2}{\td^2} \left[ - \frac{\ta''}{\ta} - \frac{\ta'}{\ta} \left(3\frac{\ta'}{\ta}
- \frac{\ta'}{\ta} \right) - \frac{1}{6} \alpha_0^2 \frac{\delta^2}{\varepsilon^2} \frac{\td^2}{\tV^2}
- \sqrt{2} \alpha_0 \frac{\delta}{\varepsilon} \frac{\td}{\tV} \left( \delta(y) - \delta(y - \pi \lambda)
\right) \right] \nonumber \\
= \frac{\delta^2}{n^2} \left[ \frac{\dot{\beta}}{\beta} \left( \frac{\dot{n}}{n} -
\frac{\dot{\alpha}}{\alpha} - \frac{\dot{\gamma}}{\gamma} - \frac{\dot{\delta}}{\delta} \right)
- \frac{\ddot \beta}\beta -\frac{n^2}{2\alpha^2\beta^2\gamma^2}\left( \left(
\alpha^2-\gamma^2\right)^2-\beta^4\right) \right] , \\
\frac{\ta^2}{\td^2} \left[ - \frac{\ta''}{\ta} - \frac{\ta'}{\ta} \left(3\frac{\ta'}{\ta}
- \frac{\ta'}{\ta} \right) - \frac{1}{6} \alpha_0^2 \frac{\delta^2}{\varepsilon^2} \frac{\td^2}{\tV^2}
- \sqrt{2} \alpha_0 \frac{\delta}{\varepsilon} \frac{\td}{\tV} \left( \delta(y) - \delta(y - \pi \lambda)
\right) \right] \nonumber \\
= \frac{\delta^2}{n^2} \left[ \frac{\dot{\gamma}}{\gamma} \left( \frac{\dot{n}}{n} -
\frac{\dot{\alpha}}{\alpha} - \frac{\dot{\beta}}{\beta} - \frac{\dot{\delta}}{\delta} \right)
- \frac{\ddot \gamma}\gamma -\frac{n^2}{2\alpha^2\beta^2\gamma^2}\left( \left(
\alpha^2-\beta^2\right)^2-\gamma^4\right) \right]  ,\\
\frac{\ta^2}{\td^2} \left[ 4 \left( \frac{\ta'}{\ta}
\frac{\td'}{\td} - \frac{\ta''}{\ta} \right) - \frac{1}{6}
\alpha_0^2 \frac{\delta^2}{\varepsilon^2} \frac{\td^2}{\tV^2} -
\frac{1}{2} \frac{\tV'^2}{\tV^2} \right]
= \frac{\delta^2}{n^2} \left[ \frac{\dot{\delta}}{\delta}
\left( \frac{\dot{n}}{n} - \frac{\dot{\alpha}}{\alpha} - \frac{\dot{\beta}}{\beta} -
\frac{\dot{\gamma}}{\gamma} \right) - \frac{\ddot \delta}\delta
\right] .
\label{fe}
\end{eqnarray}
The equation of motion (2) for the scalar field $V$ is
\begin{eqnarray}
\frac{\ta^2}{\td^2} \left[ 4 \frac{\ta'}{\ta} \frac{\tV'}{\tV} -
\frac{\td'}{\td} \frac{\tV'}{\tV} - \frac{\tV''}{\tV} +
\frac{\tV'^2}{\tV^2} \right]
= \frac{\delta^2}{n^2} \left[ \frac{\dot{\varepsilon}}{\varepsilon}
\left( \frac{\dot{\alpha}}{\alpha} + \frac{\dot{\beta}}{\beta} + \frac{\dot{\gamma}}{\gamma}
+ \frac{\dot{\delta}}{\delta} - \frac{\dot{n}}{n} \right) +
\frac{\ddot{\varepsilon}}{\varepsilon} -
\frac{\dot{\varepsilon}^2}{\varepsilon^2} \right]  .
\label{phieom}
\end{eqnarray}

In order to separate Equations (10)-(15) one has to make a choice $\delta = \varepsilon$
and if one additionally choose a gauge in the form $n = 1$ as in
Ref. \cite{lukas} one gets the following set of time-dependent field equations
(note that equations (14) and (15) become identical so we
reduce the number of equations to five \footnote{This, of course, reduces the generality of our
discussion in the context of \cite{halp} since we identify an extra dimension with a scalar
field.})
\begin{eqnarray}
\frac{\ddot \alpha}\alpha+\frac{\ddot \beta}\beta+\frac{\ddot \gamma}\gamma + \frac{\ddot \delta}
\delta = - \frac{1}{2} \frac{\dot{\delta^2}}{\delta^2}, \\
\frac{\ddot \alpha}\alpha+\frac{\dot \alpha}\alpha \left( \frac{\dot \beta}\beta +
\frac{\dot \gamma}\gamma + \frac{\dot \delta}\delta \right) =
\frac 1{2\alpha^2\beta^2\gamma^2}\left[ \left(
\beta^2-\gamma^2\right)^2-\alpha^4\right] , \\
\frac{\ddot \beta}\beta+\frac{\dot \beta}\beta \left( \frac{\dot \alpha}\alpha +
\frac{\dot \gamma}\gamma + \frac{\dot \delta}\delta \right) =
\frac 1{2\alpha^2\beta^2\gamma^2}\left[ \left(
\alpha^2-\gamma^2\right)^2-\beta^4\right] , \\
\frac{\ddot \gamma}\gamma+\frac{\dot \gamma}\gamma \left( \frac{\dot \alpha}\alpha +
\frac{\dot \beta}\beta + \frac{\dot \gamma}\gamma \right) =
\frac 1{2\alpha^2\beta^2\gamma^2}\left[ \left(
\alpha^2-\beta^2\right)^2-\gamma^4\right]  ,\\
\frac{\ddot \delta}\delta + \frac{\dot \delta}\delta \left(
\frac{\dot \alpha}\alpha + \frac{\dot \beta}\beta + \frac{\dot
\gamma} \gamma \right) = 0   ,
\label{fet}
\end{eqnarray}
which, except for the right-hand side of Eq.(16), is the same set as the
set of equations (3.7)-(3.11) of \cite{bd2} for bosonic low-energy-effective-action
cosmology in string frame, provided we take dilaton field $\phi$ as
defined in \cite{bd2} to be equal to $(-\ln{\delta})$ and also neglect
axion (i.e. take $A=0$ in Ref. \cite{bd2}).

A new time coordinate is introduced to simplify the field equations by
(compare Eq. (3.12) of \cite{bd2})
\begin{equation}
d\eta =\frac{d\tau}{\alpha\beta\gamma\delta}  .
\end{equation}
From now on we will use the notation $(...),_{\eta} = d/d\eta$. To
further simplify the equations we additionally define
\begin{equation}
\tilde{\alpha} = \ln{\alpha}\hspace{2.0cm}\tilde{\beta} = \ln{\beta}
\hspace{2.0cm}\tilde{\gamma} = \ln{\gamma }\hspace{2.0cm}
\tilde{\delta} = \ln{\delta} ,
\end{equation}
so that the set of equations (16)-(20) reads as
\begin{eqnarray}
\left(\tilde{\alpha} + \tilde{\beta} +\tilde{\gamma} + \tilde{\delta} \right) _{, \eta \eta }
+ \frac{1}{2} \tilde{\delta}_{,\eta}^2
&=& 2\left(\tilde{\alpha}_{,\eta}\tilde{\beta}_{,\eta} +
\tilde{\alpha}_{,\eta}\tilde{\gamma}_{,\eta} + \tilde{\beta}_{,\eta}\tilde{\gamma}_{,\eta}\right)
+ 2\left(\tilde{\alpha}_{,\eta} + \tilde{\beta}_{,\eta } + \tilde{\gamma}_{,\eta}\right)
\tilde{\delta}{,\eta }, \\
2\tilde{\alpha}_{,\eta \eta } &=& \left[ \left(
\beta^2-\gamma^2 \right)^2-\alpha^4 \right]\delta^2, \\
2\tilde{\beta}_{,\eta \eta} &=&\left[ \left(
\alpha^2-\gamma^2\right)^2-\beta^4 \right]\delta^2, \\
2\tilde{\gamma}_{, \eta \eta } &=&\left[ \left(
\beta^2-\gamma^2\right)^2-\alpha^4 \right]\delta^2,  \\
\tilde{\delta}_{'\eta\eta} &=& 0 .
\end{eqnarray}

These equations are the same as pre-big-bang cosmology Mixmaster equations in string frame
(3.19)-(3.22) of Ref. \cite{bd2} if we take $\tilde{\delta} = - \phi = - M \eta + $const
or as 5-dimensional vacuum Mixmaster equations (31)-(32) of Ref.
\cite{halp}.

Now we consider suitable initial conditions expressed in terms of the Kasner
parameters and discuss the general behaviour of Bianchi type IX Ho\v{r}ava-Witten cosmology
on the approach to singularity.

The Kasner solutions are obtained as approximate solutions of the
equations (16)-(20) when the right-hand sides (describing the curvature anisotropies)
are neglected. In terms of $\tau$-time, they are
\begin{eqnarray}
\alpha &=& \alpha_0 \tau^{p_1},  \nonumber \\
\beta &=& \beta_0 \tau^{p_2}, \\
\gamma &=& \gamma_0 \tau^{p_3}, \nonumber \\
\delta &=& \delta_0 \tau^{p_4}, \nonumber
\end{eqnarray}
while
\begin{equation}
\tilde{\delta} = - \ln {\delta_0} - p_4 \ln{\tau}  .
\end{equation}

{}From (23)-(27) we have the following algebraic conditions for the Kasner indices, $p_i:$
\begin{equation}
p_1 + p_2 + p_3 + p_4 = 1,
\end{equation}
and
\begin{equation}
p_1^2 + p_2^2 + p_3^2 + \frac{3}{2}p_4^2 = 1 .
\end{equation}

This, in particular, proves that the isotropic Friedmann case as obtained by
Lukas et al. \cite{lukas} is given by
\bea
p_1 = p_2 = p_3 = p_{\mp} = \frac{3}{11} \mp
\frac{4}{11\sqrt{3}}   ,\\
p_4 = q_{\pm} = \frac{2}{11} \pm \frac{4\sqrt{3}}{11}   .
\eea

The meaning of such isotropic solutions was discussed in Ref. \cite{lukas} (note that numerically
$p_+ = 0.48$, $p_- = 0.06$ and $q_- = -0.45$, $q_+ = 0.81$). In fact, there are two branches each one for negative
and positive values of time coordinate $\tau$ (negative values of time can be achieved by taking
$-\tau$ instead of $+\tau$ in (28), or, simply by taking the modulus).
If $\tau < 0$ one has $(-)$ branch and if $\tau > 0$ one has $(+)$
branch. For $(-)$ branch both the vorldvolume (of 3-dimensional space) and the
orbifold contract for $p_-$ and $q_+$ while the worldvolume contracts and the
orbifold expands (superinflationary) for $p_+$ and $q_-$. For $(+)$
branch the worldvolume and the orbifold expand for $p_-$ and $q_+$
while the wordvolume expands and the orbifold contracts for $p_+$
and $q_-$.

Notice that the conditions (30)-(31) are different from the
conditions which emerge in pre-big-bang cosmology where the role
of the fifth coordinate is played by the dilaton (see Eqs.(3.50)-(3.51) of \cite{bd2}).
They are also different from
Mixmaster Kaluza-Klein five-dimensional models where the
homogeneity group acts on four-dimensional hypersurfaces of
constant time (see Eq.(33) of \cite{halp}). The reason for that is
simply the fact the fifth coordinate in Ho\v{r}ava-Witten
cosmology is an orbifold. The isotropization of the models under
consideration means that the Kasner indices reach the values
(32)-(33). Finally, the isotropization of 5-dimensional Kaluza-Klein
models in supergravity as first considered by Chodos and Detweiler
\cite{chodos} would require the different values of the Kasner
indices namely $p_1 = p_2 = p_3 = - p_4 = 1/2$ which fulfill the
conditions (30) and (31) without a factor $3/2$ in front of $p_4$
in (31).

Having given the conditions (30)-(31), one can express the indices
$p_2$ and $p_3$ by using $p_1$ and $p_4$, i.e.,
\bea
p_2 = \frac{1}{2} \left[ \left( 1 - p_1 - p_4 \right) - \sqrt{ - 3p_1^2 +
2p_1 \left(1 - p_4 \right) + 1 + 2p_4 \left(1 - 2p_4 \right)} \right]  \nonumber,\\
p_3 = \frac{1}{2} \left[ \left( 1 - p_1 - p_4 \right) + \sqrt{ - 3p_1^2 +
2p_1 \left(1 - p_4 \right) + 1 + 2p_4 \left(1 - 2p_4 \right)}
\right].
\eea
Since the expression under the square root in (34) should be nonnegative, one can extract
the restriction on the permissible values of $p_4$ which is
\be
q_{-} \leq p_4 \leq q_{+}   .
\ee

Some particular choices are of interest. If one takes $p_4 = 0$
one recovers vacuum general relativity limit with Kasner indices $-1/3 \leq p_1 \leq 0,
0 \leq p_2 \leq 2/3, 2/3 \leq p_3 \leq 1$. The range dividing case is
for $p_4 = 2/11$ with the following ordering of the Kasner indices
\bea
\frac{3}{11} - \frac{4}{11\sqrt{3}} \sqrt{11} \leq p_1 \leq \frac{3}{11} - \frac{2}{11\sqrt{3}}
\sqrt{11} \nonumber ,\\
\frac{3}{11} - \frac{2}{11\sqrt{3}} \sqrt{11} \leq p_2 \leq \frac{3}{11} + \frac{2}{11\sqrt{3}}
\sqrt{11} ,\\
\frac{3}{11} + \frac{2}{11\sqrt{3}} \sqrt{11} \leq p_3 \leq \frac{3}{11} + \frac{4}{11\sqrt{3}}
\sqrt{11} \nonumber .
\eea

However, we are interested in knowing whether the curvature terms on the right-hand side of
the field equations (24)-(26) really increase as $\eta \to - \infty$ ($\tau \to 0$
- approach to singularity for (+) branch) since from (21) and (28) we get
\be
\eta = \eta_0 + \ln{\tau} ,
\ee
and $\eta_0=$ const.
This would require either $\alpha^4\delta^2, \beta^4\delta^2$, or
$\gamma^4\delta^2$ to increase if the transition to another Kasner epoch is to
occur \cite{halp,ishihara}. Since
\bea
\alpha^4\delta^2 \propto \tau^{(2p_1 + p_4)} = \tau^{(1 + p_1 - p_2 - p_3)}   \nonumber,\\
\beta^4\delta^2 \propto \tau^{(2p_2 + p_4)} = \tau^{(1 + p_2 - p_3 - p_1)}   ,\\
\gamma^4\delta^2 \propto \tau^{(2p_3 + p_4)} = \tau^{(1 + p_3 - p_1 - p_2)}   \nonumber,
\eea
we need one of the following three conditions to be fulfilled
\bea
2 p_1 + p_4 & = & 1 + p_1 - p_2 - p_3  <  0   \nonumber,\\
2 p_2 + p_4 & = & 1 + p_2 - p_3 - p_1  <  0   ,\\
2 p_3 + p_4 & = & 1 + p_3 - p_1 - p_2  <  0   \nonumber.
\eea

{}From Kasner conditions (30)-(31) we are free to choose only
two parameters so that we can write
\bea
p_4 &=& 1 - p_1 - p_2 - p_3  , \nonumber \\
p_3 &=& \frac{3}{5} \left(1 - p_1 - p_2 \right) \pm \frac{2}{5} \left[ -
4\left(p_1^2 + p_2^2 \right) - 3p_1p_2 + 3p_1 + 3p_2 + 1
\right]^{\frac{1}{2}}   ,
\eea
which gives also the condition for $p_3$ to be real as follows
\be
1 - 4\left(p_1^2 + p_2^2 \right) - 3p_1p_2 + 3p_1 + 3p_2 \geq 0.
\ee
In (40) we take the $+$ sign for $p_4 < 2/11$ and the $-$ sign for
$p_4 > 2/11$. The conditions (39) become (compare \cite{halpern})
\bea
p_1^2 + 4p_2^2 - p_1p_2 + p_2 - p_1 & < & 0  , \nonumber\\
4p_1^2 + p_2^2 - p_1p_2 + p_1 - p_2& < & 0   ,\\
4\left(p_1^2 + p_2^2\right) + 7p_1p_2 - 7\left(p_1 + p_2\right) + 3& > & 0 .   \nonumber
\eea

One should remind here that for (-) branch we have to take $(-\tau)$ in (28), (29) and (38)
which leads to the same conditions (42).

The plot of the conditions (41)-(42) is given in Fig.1. The chaotic oscillations (Kasner-to-Kasner
transitions) can start in any region except the narrow range surrounding the isotropic
points $p_1 = p_2 = p_+ (q = q_-)$ and $p_1 = p_2 = p_- (q = q_+)$. However, such chaotic
oscillations would continue indefinitely provided there
were no such regions at all (this is the case of vacuum general relativity, for example).
Here, once the Kasner parameters fall into the region surrounding the isotropic
Friedmann $p_+ $ or $p_-$ solutions of Lukas et al. \cite{lukas} the
chaotic oscillations cease so that there is no chaos in such Ho\v{r}ava-Witten cosmologies.

{\bf Acknowledgments}

This work was supported by the Polish Research Committee (KBN) grant No 2 PO3B 105 16.

\begin{figure}[t]
\centering
\leavevmode\epsfysize=14cm \epsfbox{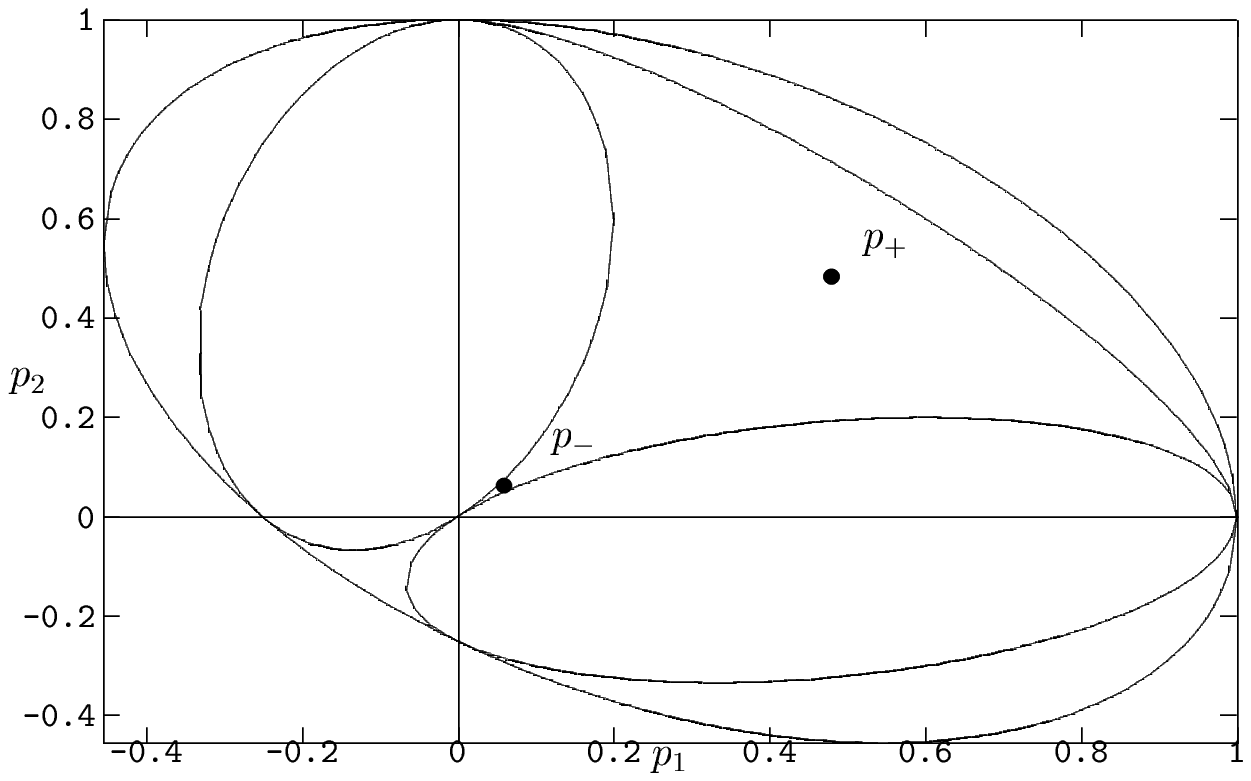}\\
\caption[]
{The range of Kasner indices $p_1$ and $p_2$ which fulfill the conditions (41)-(42).
The appearance of the isotropic Friedmann cases at $p_1 = p_2 = p_-$  and $p_1 = p_2 = p_+$
prevents chaotic oscillations in the shaded region that surrounds them. For the values of Kasner
indices $p_1$ and $p_2$ from that region chaotic oscillations are not possible to begin. On the
other hand, even if after some number of oscillations from one
Kasner epoch to the other, the values of $p_1$ and $p_2$ will fall into that region, the chaotic
oscillations of the scale factors stop which reflects nonchaotic behaviour of such Bianchi IX
Ho\v{r}ava-Witten cosmologies.}
\label{fig1}
\end{figure}


\begin{thebibliography}{99}

\bibitem{superjim} J.E. Lidsey, D.W. Wands and E.J. Copeland,
                   Superstring Cosmology, e-print hep-th/9909061.

\bibitem{superinf} G. Veneziano, Phys. Lett. B{\bf 406}, 287
                   (1991); M. Gasperini and G. Veneziano,
                   Astroparticle Phys. {\bf 1}, 317 (1991).

\bibitem{fradcall} E.S. Fradkin and A.A. Tseytlin, Nucl. Phys. B{\bf 261},
               1 (1985);\\
               C.G. Callan, D. Friedan, E.J. Martinec and M.J. Perry, Nucl.
               Phys B{\bf 262}, 593 (1985).

\bibitem{hw} P. Ho\v{r}ava and E.Witten, Nucl. Phys. B{\bf 460} (1996)
             506; {\it ibid} B{\bf 475} (1996) 94.

\bibitem{dimo} N. Arkani-Hamed, S. Dimopoulos and G. Dvali, Phys. Lett. B
               {\bf 249} (1998) 263; Phys. Rev. D{\bf 59} (1999)
               086004.

\bibitem{kalo} N. Kaloper, Bent Domain Walls as Braneworlds, e-print
               hep-th/9905210; Phys. Rev. D to appear.

\bibitem{extradim} J.M. Cline, C. Grojean and G. Servant, Cosmological
                   Expansion in the Presence of an Extra Dimension, e-print
                   hep-ph/9906523; Phys. Rev. Lett. to appear.

\bibitem{kam} K.A. Meissner and M. Olechowski, Towards
              Supersymmetric Cosmology in M-theory, e-print
              hep-th/9910161.

\bibitem{extratime} G. Dvali, G. Gabadadze and G. Senjanovi\'c,
                  Constraints on Extra Time Dimensions, e-print hep-ph/9910207.

\bibitem{lukas} A. Lukas, B.A. Ovrut and D. Waldram, Phys. Rev. D{\bf 60} 086001 (1999).

\bibitem{reall} H.S. Reall, Phys. Rev. D{\bf 59}, 103506 (1999).

\bibitem{bd2} J.D. Barrow and M.P. D\c{a}browski, Phys. Rev. D{\bf 55},
                7204 (1998).

\bibitem{bkl} V.A. Belinskii, E.M. Lifshitz and I.M. Khalatnikov, Sov. Phys.
         Uspekhi {\bf 102}, 745 (1971);\\
         J.D. Barrow, Phys. Rep. {\bf 85}, 1 (1982).

\bibitem{belkhal} V.A. Belinskii and I.M. Khalatnikov, Sov. Phys. JETP {\bf 36},
        591 (1973).

\bibitem{john85} J.D. Barrow and J. Stein-Schabes, Phys. Rev. D{\bf32}, 1595
        (1985).

\bibitem{ishihara} H. Ishihara, Prog. Theor. Phys. {\bf 74}, 354 (1986).

\bibitem{halp} P. Halpern, Phys. Rev. D{\bf 33}, 354 (1986).


\bibitem{chodos} A. Chodos and S. Detweiler, Phys. Rev. D{\bf 21}
                (1980) 2167.

\end{thebibliography}
\end{document}